\documentclass[a4paper,fleqn,usenatbib]{mnras}

\usepackage[T1]{fontenc}
\usepackage{ae,aecompl}

\usepackage{graphicx}	
\usepackage{amsmath}	
\usepackage{amssymb}	
\usepackage{lipsum}
\usepackage{verbatim}
\usepackage{subfigure}

\usepackage{xcolor}

\newcommand{\chandra}{{\it CHANDRA}}

\newcommand{\einstein}{{\it EINSTEIN}}

\newcommand{\xmm}{{\it XMM-NEWTON}}

\newcommand{\swift}{{\it Swift}}

\newcommand{\nustar}{\textit{NuSTAR}}

\newcommand{\ms}{$M_{\odot}$}

\newcommand{\lumcgs}{ergs~s$^{-1}$}

\title[Super-Eddington accretion onto NGC 4190 ULX1]{Super-Eddington accretion onto a stellar mass ultraluminous X-ray source NGC 4190 ULX1}

\author[Ghosh et al.]{
T.Ghosh, \thanks{E-mail: tanuman@rri.res.in}
V.Rana  \thanks{E-mail: vrana@rri.res.in}
\\
Raman Research Institute, C.V.Raman Avenue, Sadashivanagar, Bangalore-560080, India\\
}

\date{Accepted XXX. Received YYY; in original form ZZZ}

\pubyear{2020}

\begin{document}
\label{firstpage}
\pagerange{\pageref{firstpage}--\pageref{lastpage}}
\maketitle

\begin{abstract}
We present the results of high-quality \xmm\ observations of a ULX in the galaxy NGC 4190. The detection of spectral cutoff in NGC 4190 ULX1 spectra rules out the interpretation of the ULX to be in a standard low/hard canonical accretion state. We report that the high quality EPIC spectra can be better described by broad thermal component, such as a slim disk. In addition we found long term spectral and flux variability in the source using several \xmm\ and \swift\ data. A clear anti-correlation between flux and power-law photon index is found which further confirms the unusual spectral state evolution of the ULX. Spectral properties of the ULX suggest that the source is in a broadened disk state with luminosities ($\approx (3-10) \times  10^{39}$ \lumcgs) falling in the ultraluminous regime.  The positive Luminosity-temperature relation further suggests that the multi color disk model follows the $L \propto T^4$ relation which is expected for a black body disk emission from a constant area and the slim disk model seems to favour $L \propto T^2$ relation consistent with an advection dominated disk emission . From the broadened disk like spectral feature at such luminosity, we estimated the upper limit of the mass of the central compact object from the inner disk radius and found that the ULX hosts a stellar mass black hole.
\end{abstract}




\section{Introduction}

Ultraluminous X-ray sources (ULXs) are one of the most fascinating albeit one of the least apprehended sources in the field of X-ray astronomy. These are extragalactic off-nuclear point sources with X-ray luminosity ($L_x> 10^{39}$ \lumcgs) exceeding the typical isotropic Eddington limit of a stellar remnant black hole ($M_{BH} \sim 10$ \ms). Since the first detection of these sources by \einstein\ telescope \citep{Fabbiano1}, study of ULXs have become highly intriguing because of their distinct nature from the well studied Galactic X-ray binary (XRB) sources. Apart from their high X-ray luminosity, many of these sources have shown long and short-term variability like Galactic XRBs which further confirm them to be accreting binaries \citep{Miller1}. 

It is still a mystery what is the power house of such high luminosity in these sources. Possible scenarios are intermediate mass black holes (IMBHs) emitting with a sub-Eddington process \citep{Colbert1}, a stellar mass compact object emitting with super-Eddington accretion \citep{Begelman1, Ebisawa1}, highly relativistic beamed emission\citep{Kording1} or geometrically beamed emission \citep{King1} from which such high luminosity is generated in ULXs.

Recent observations with high quality data from \xmm\ , \chandra\ , \nustar\ have established the emission mechanism for majority of ULX sources as super-Eddington emission from a stellar mass X-ray binary \citep{Bachetti1, Walton1, Walton2, Walton3, Walton4, Rana1, Mukherjee1, Furst1}. In fact, discovery of pulsating ULXs (PULXs) \citep{Bachetti2, Furst2, Israel1, Israel2, Carpano1, Sathyaprakash1, Rodriguez1}, cyclotron line \citep{Brightman1} in ULX spectra and potential bi-modal flux distribution \citep{Earnshaw1} further confirmed the notion of super-Eddington emission mechanism from neutron stars. The spectral nature of these ULX sources are unusual when compared to known Galactic XRBs in {\it hard} and {\it soft} state and hence referred to as the ``ultraluminous state" \citep{Roberts1, Gladstone1}. This spectral state is best described as a manifestation of super-Eddington accretion process.

Spectra from several ULXs have been found with spectral curvature around $\sim 3-10$ keV \citep{Kaaret1, Bachetti1, Walton1, Rana1} unlike Galactic XRBs which have curvature at much higher energies. Various possible physical scenarios have been invoked to understand the origin of such spectral curvature, these are, comptonization from optically thick, cold corona \citep{Gladstone1}, a relativistically smeared iron features in a blurred reflection of coronal emission from an accretion disk \citep{Caballero1}, a modified ``slim" accretion disk or the hot inner regions of the disk highly distorted by advection, turbulence, self-heating and spin (\citep{Pintore1} and references therein).

Broadband X-ray spectra of some bright ULXs \citep{Kaaret1} proved that a slim disk model is highly preferred over a thin Keplerian disk and a high energy comptonization component is required in the hard spectral tail ($\geq 10$ keV). In case of super-Eddington emission, outward radiation pressure increases the scale height of the inner most portion of the accretion disk leading to a funnel like ``slim" disk structure \citep{West1} which creates a radiatively driven wind and a soft excess from the photosphere of the wind ejected from the spherization radius of the disk down to its innermost region \citep{Shakura1}. The outflowing material obscures the hot inner disk and give a cool but optically thick comptonized corona which has a signature in hard spectral excess.

In this paper we explore a nearby ($D=3 $ Mpc), bright ($\sim 3-10 \times 10^{39}$ \lumcgs in $0.3-10.0$ keV) and variable ULX source CXO J121345.2+363754 (hereafter NGC 4190 ULX1) in a low surface brightness galaxy NGC 4190. We perform detail spectral and timing analysis of all archival \xmm\ observations and studied the transient nature of the source. We have also analysed  good signal-to-noise ratio \swift\ observations to study long term spectral variability.

The paper is organized as follows. In Sect. ~\ref{sec:obs} we describe the analysis of \xmm\ and \swift\ data. Detail timing and spectral analyses and variability studies are presented in Sect. ~\ref{subsec:timing}, ~\ref{subsec:spectra} and \ref{subsec:variable} respectively. In Sect. ~\ref{sec:discuss} and ~\ref{sec:conclusion}, we discuss and conclude our results.

\section{Observations and Data Reduction}
NGC 4190 ULX1 was observed three times with \xmm\ \citep{Jansen1} and six times with \swift\ \citep{Gehrels1} (See table \ref{tab:log_table} for observation log). In this work we mainly focus on the \xmm\ -EPIC data for detailed timing and spectral analysis. For long-term spectral variability study, we use both \xmm\ and \swift\ -XRT data.

\label{sec:obs}

\subsection{\it XMM-NEWTON}
The galaxy NGC 4190 was observed three times by \xmm\ in 2010.
We carried out the data reduction using the \xmm\ Science Analysis System (SAS v18.0.0). Calibrated event lists for the EPIC PN and MOS detectors are produced using the SAS tools {\texttt{epproc}} and {\texttt{emproc}} respectively. To remove background flaring contribution and generate clean event files, we used {\texttt{espfilt}} task.

Pileup was evaluated using {\texttt{epatplot}} tool and none of the observations were affected by pileup. The filtered cleaned events are used to generate source and background spectra for circular regions with $30 ^{\prime\prime}$ and $60^{\prime \prime}$ radii respectively on the same CCD using {\texttt{evselect}} task. RMFs and ARFs are generated using {\texttt{rmfgen}} and {\texttt{arfgen}} tools. The X-ray spectra are grouped to have a minimum of 20 counts per energy bin. The {\texttt{evselect}} task is also used to extract light curves from the same source and background regions taking single and double events for PN (PATTERN<=4) and singles, doubles, triples and quadruples events for MOS (PATTERN<=12) in $0.3-10.0$ keV energy range. To generate background corrected source light curve, {\texttt{epiclccorr}} tool is used. We found that the first \xmm\ observation (Epoch 1) is highly contaminated with background flaring activity, hence background corrected PN exposure is too low (103 sec) for any useful scientific analysis. So, we did not use the Epoch 1 PN data for any further analysis.

\subsection{\it Swift}
Swift observed NGC 4190 six times from 2014 to 2019. Analysis of \swift\ data was executed in $0.3-10.0$ keV energy range. The X-ray products are generated using the {\texttt{xrtpipeline}} tool that is part of FTOOLS (HEAsoft 6.27.2) software package.  Spectra are extracted using {\texttt{xselect}} tool with a $30 ^{\prime\prime}$ source region and $60 ^{\prime\prime}$ background region. The XRT spectra are grouped to have a minimum of 1 count per energy bin. Owing to the low count statistics of {\it Swift} XRT spectra, we used cash statistics ({\texttt{cstat}}) \citep{Cash1} for the spectral analysis. As mentioned previously, we used Swift data for variability study of different spectral parameters.

\begin{table*}
	\centering
	\caption{X-ray observations list of NGC 4190 ULX1.}
	\label{tab:log_table}
	\begin{tabular}{lccrcr} 
		\hline
		Mission & Date & ObsID & &Exposure (ks)$^*$ &\\
		\hline
		& & & PN & MOS1 & MOS2 \\
		XMM $^\dagger$ & 2010-06-06 (Epoch 1) & 0654650101 & 0.1 & 3.7 & 4.0 \\
		XMM & 2010-06-08 (Epoch 2) & 0654650201 & 4.0 & 12.4 & 12.3\\
		XMM & 2010-11-25 (Epoch 3) & 0654650301 & 6.3 & 10.4 & 10.9 \\
		\hline
		& & &  & XRT  &  \\
		Swift & 2014-10-14 & 00084393001 & &3.4 &\\
		Swift & 2016-01-25 & 00084393002 & &0.6 &\\
		Swift  & 2017-10-14 & 00084393003 & & 0.8  &\\
		Swift & 2019-01-23 & 00084393004 & & 0.5  &\\
		Swift & 2019-03-05 & 00084393005 & &0.9  &\\
		Swift & 2019-11-28 & 00084393006 &  &1.4  &\\
		\hline
	\end{tabular}
	
	$*$ The exposure mentioned here is the background flare corrected cleaned exposure for \xmm\ observation. \\
	$\dagger$ This \xmm\ observation is affected by high flaring background, hence PN data is not used due to its low cleaned exposure. \\
\end{table*}

\section{Results}

\subsection{Timing Analysis}
\label{subsec:timing}
\begin{figure*}
\centering

    \includegraphics[width=\textwidth]{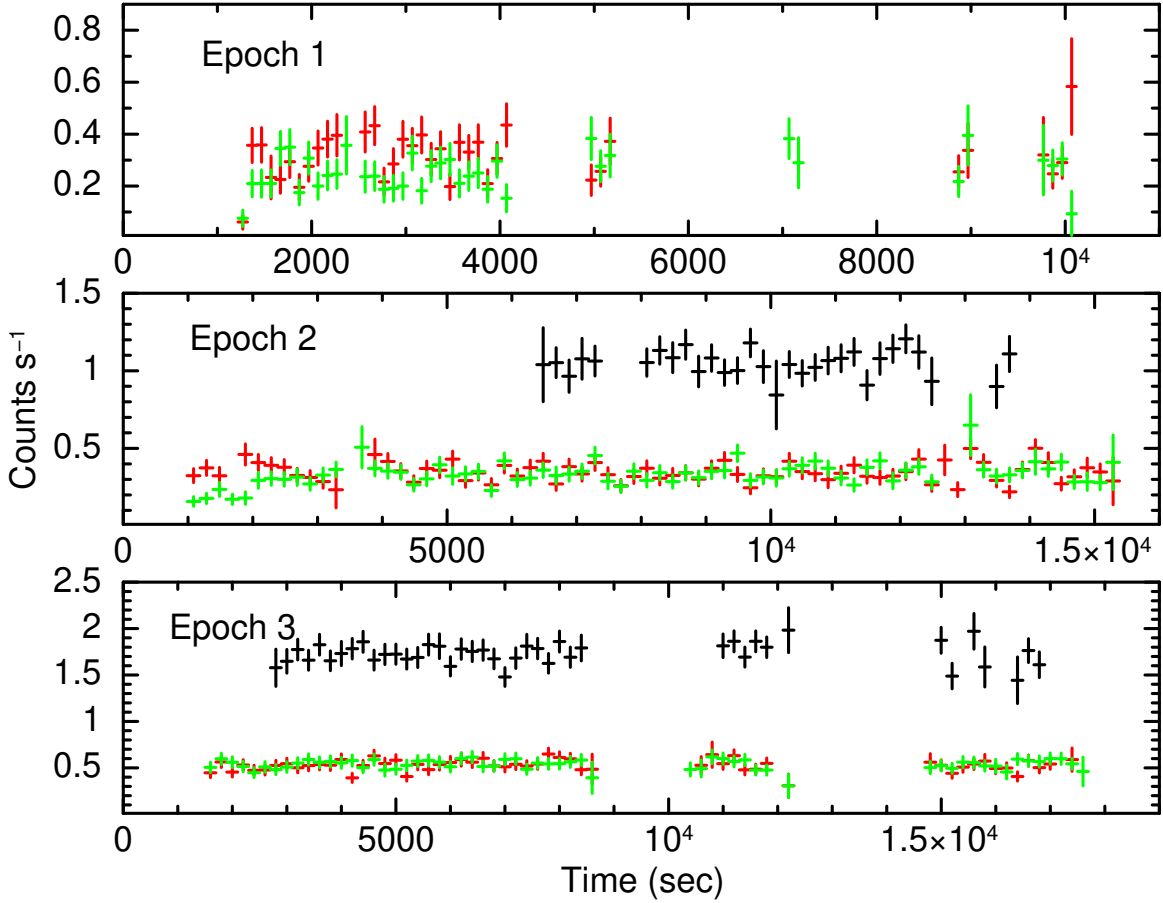}
    
    \caption{The \xmm\ light curves for three different epochs are shown in three panels. The source is steady within the exposure of each observation, however, average count rates does vary between different epochs which signifies long-term variability of the source. Epoch 1 is highly affected with flaring, so PN observation is not shown in the plot due to its low count rate. Light curves have been rebinned for visual purpose. Black, red and green represent PN, MOS1 and MOS2 light curves for all observations.}
    \label{fig:lightcurve}
\end{figure*}

The X-ray light curves from three epochs of \xmm\ observations are shown in Figure~\ref{fig:lightcurve}. The black, red and green coloured data points represents PN, MOS1 and MOS2 instruments respectively and three panels correspond to three epochs of observations as labeled. Visual inspection of these light curves indicates that the source NGC 4190 ULX1 is in constant flux state within the observation time. EPIC-MOS1 and MOS2 each has count rates of $ \sim 0.2$ cts s$^{-1}$, $\sim 0.3 $ cts s$^{-1}$, $\sim  0.5$ cts s$^{-1}$ for Epoch 1, Epoch 2 and Epoch 3 respectively. Average EPIC-PN count rates are of $\sim 1.1$ cts s$^{-1}$ for Epoch 2 and $\sim 1.7$ cts s$^{-1}$ for Epoch 3.

\begin{figure}
	\includegraphics[width=\columnwidth]{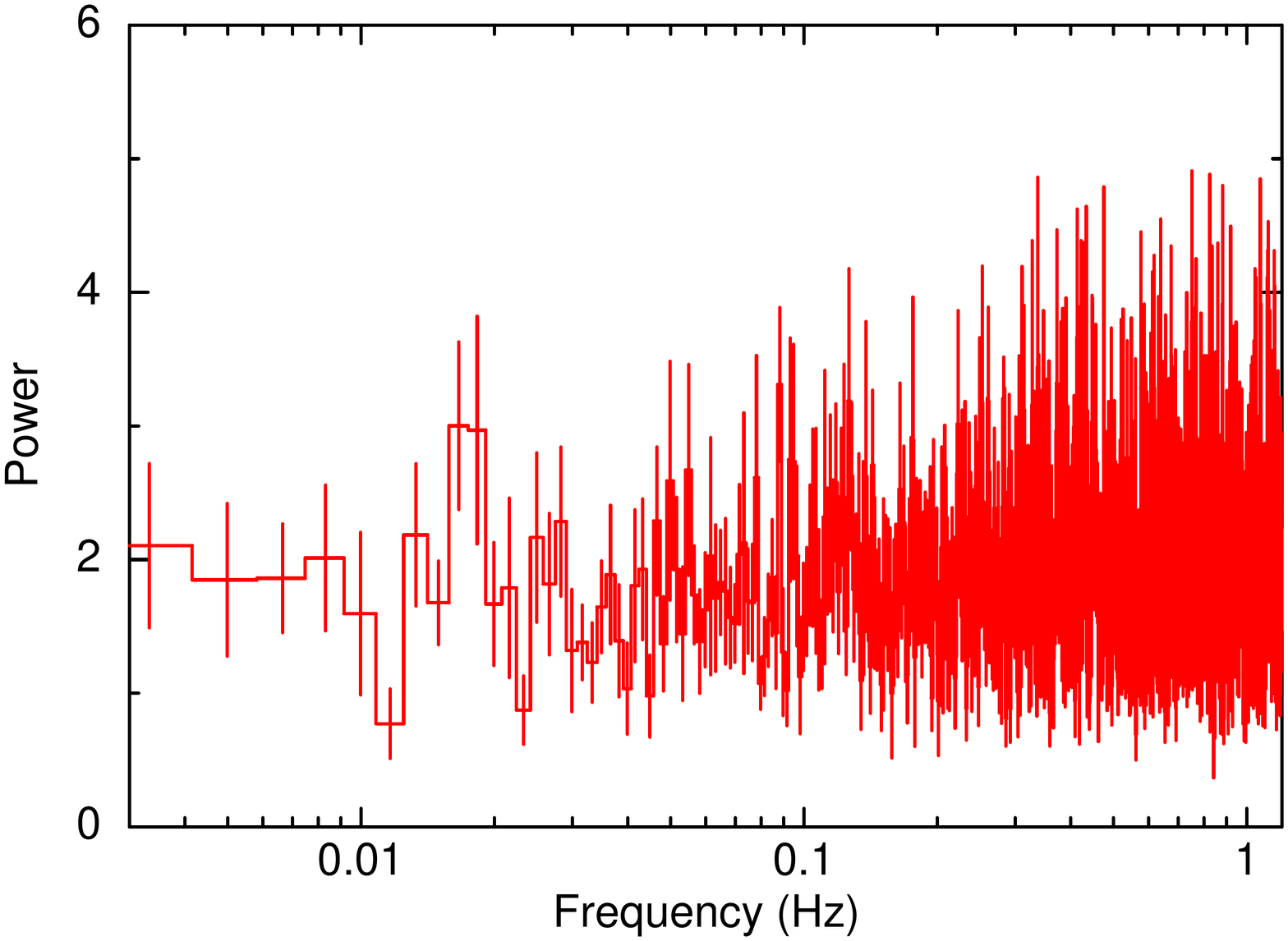}
    \caption{Power spectral density of the source from Epoch 3 (EPIC-PN) data which shows no significant variability on top of white noise.}
    \label{fig:powspec}
\end{figure}

Different average count rate values during these three epochs of observation clearly suggest that there is a long-term variability in X-ray light curves of the source, however we did not find any significant short-term variability or pulsation using the power spectral density (PSD) in any of these observation. We used fast folding $Z^2_n$ algorithm and epoch folding search with HENDRICS \citep{hendrics} tool to detect pulsation after correcting for the spin up/spin down rate of the compact object, but did not find any significant pulsations. This most likely suggests that the short-term variability is probably dominated by the white noise in the data. Since, all observations give similar PSD and lack of any significant feature, in Figure ~\ref{fig:powspec} we show the PSD for Epoch 3 (EPIC-PN) only as a representative of all observations.

\subsection{Spectral Analysis}
\label{subsec:spectra}
Throughout this work, we used XSPEC v12.11.0m \citep{Arnaud1} to perform detailed X-ray spectral analysis for NGC~4910 ULX1. The absorption effects due to neutral absorbers are modelled using {\texttt{tbabs}} with updated solar abundances \citep{Wilms1} and photoionization cross section \citep{Verner1}. We employed $\chi^2$ minimization for spectral model fitting and report the errors with $90\%$ confidence unless mentioned otherwise.

\begin{figure}
	\includegraphics[width=\columnwidth]{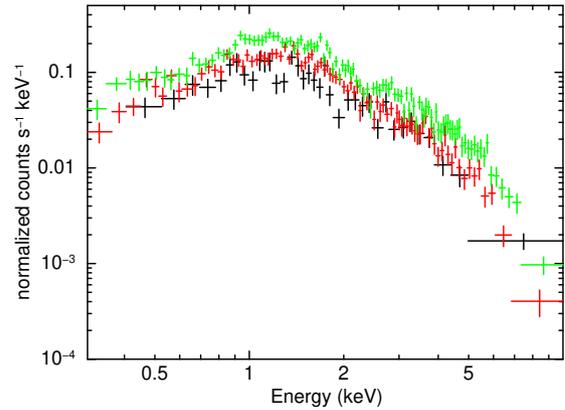}
    \caption{MOS1 Spectra for all three XMM observations. Black, red and green represent the Epoch 1, Epoch 2 and Epoch 3 respectively. Clear long-term variability is detected.}
    \label{fig:spectra}
\end{figure}

For detail spectral study, we use three available \xmm\ data. For the epoch 1 observation, we use only two MOS spectra for analysis but given their low count statistics we could only fit simple models. Figure ~\ref{fig:spectra} shows MOS1 spectra for all three observations. There is a clear indication of flux variability in ULX1. During epoch 3 (green) the source showed highest flux whereas it was at lowest flux level during epoch 1 (black). Considering longer exposures for epoch 2 and epoch 3 compared to epoch 1, we will first concentrate on results obtained from epochs 2 \& 3, then for the sake of completeness, we will discuss results from epoch 1 also. In epoch 2 and epoch 3 the analysis was performed by simultaneously fitting PN, MOS1 and MOS2 data and in the case of epoch 1 we fitted only MOS1 and MOS2 spectra simultaneously.

\begin{figure}
\centering
\subfigure[]
	{\includegraphics[width=\columnwidth]{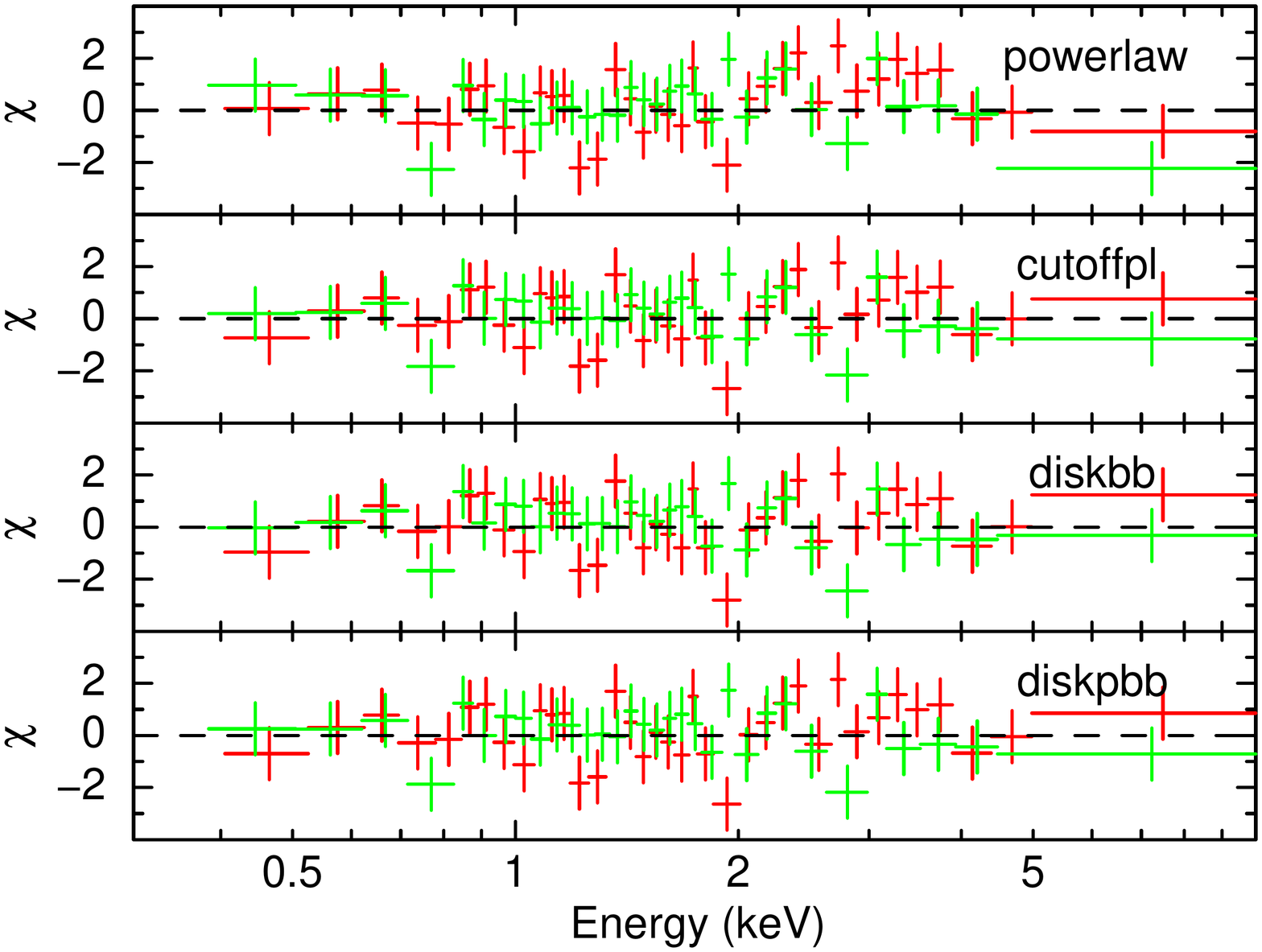}}
   
\subfigure[] {    
    \includegraphics[width=\columnwidth]{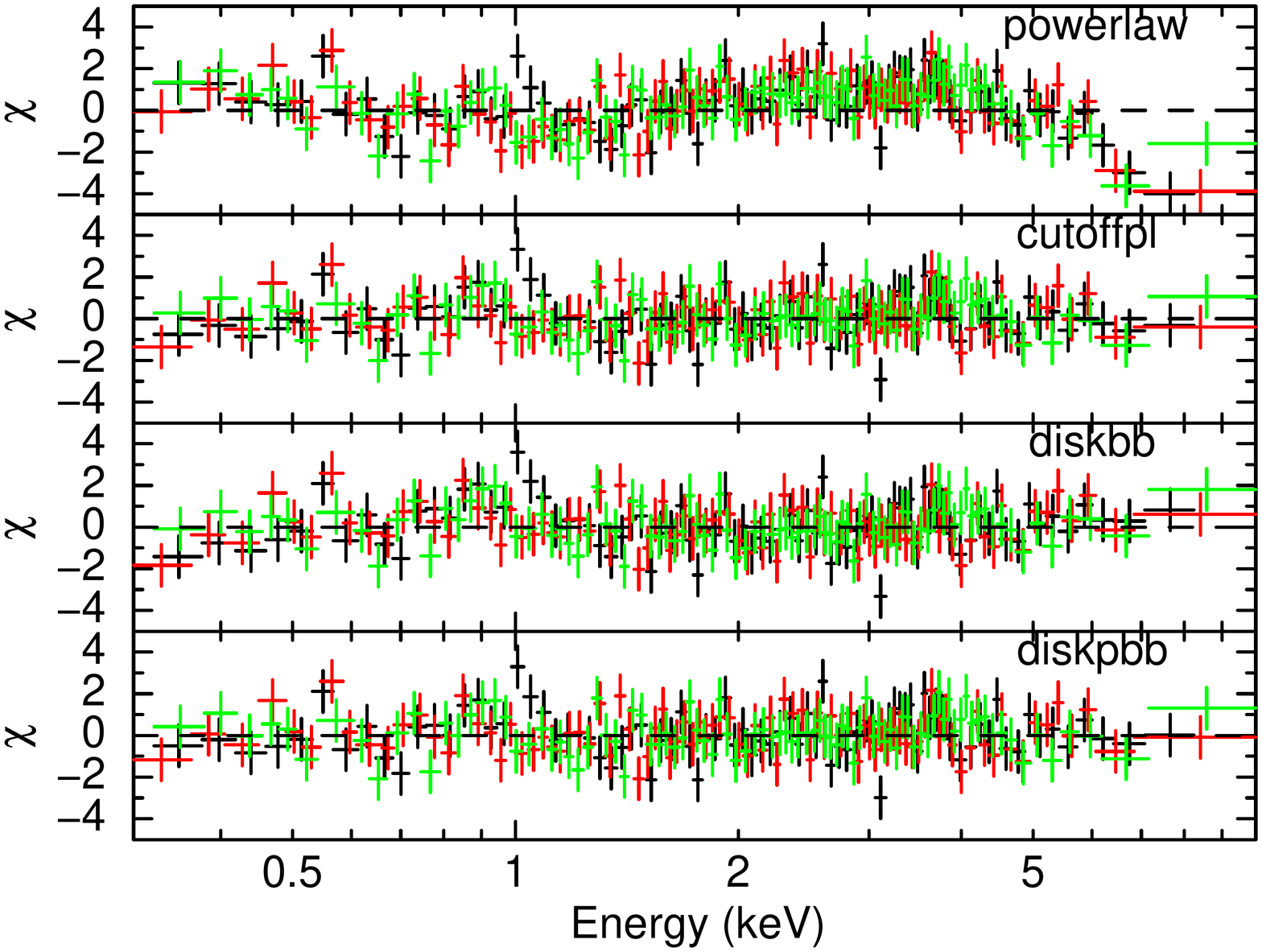}}

\subfigure[]{
    \includegraphics[width=\columnwidth]{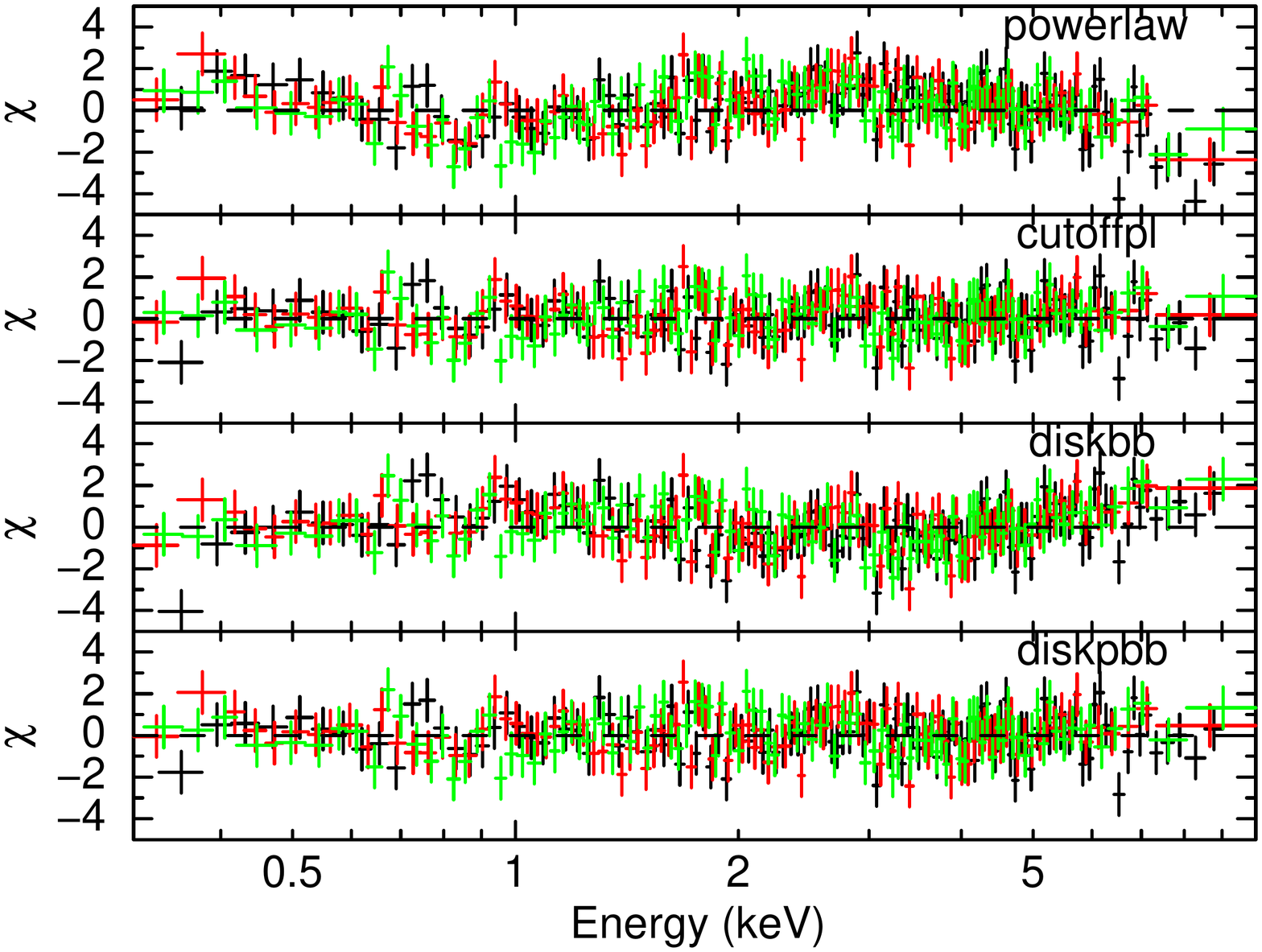}}
    
    \caption{Spectral residuals for (a) epoch 1 observation (b) epoch 2 observation (c) epoch 3 observation. Black, red and green represent PN, MOS1 and MOS2 spectra for all observations. Clear cutoff is visible when fitted with simple power-law model. Physical model of ``slim disk" accretion geometry (diskpbb) justifies this cutoff.  PN data is not used in epoch 1 due to its low on-time exposure.}
    \label{fig:main_spectral_residuals}
\end{figure}

Epoch 2 has a flaring background corrected exposure of $\sim 4$ ksec for PN, $\sim 12$ ksec for both MOS1 and MOS2. Epoch 3 has a flaring background corrected exposure of $\sim 6$ ksec for PN, $\sim 10$ ksec and $\sim 11$ ksec for MOS1 and MOS2 respectively. Thanks to high photon collecting area of PN instrument, we could obtain enough photon count statistic for spectral study. We initially fitted the EPIC-PN and EPIC-MOS spectra with absorbed power-law. It resulted in a poor statistical fit (see Table ~\ref{tab:param1}) with large residuals (see figure ~\ref{fig:main_spectral_residuals}). Since Galactic column absorption ($2.5 \times 10^{20}$ cm$^{-2}$) is very small compared to the absorption found in the spectral fitting, it is sufficient to use a single neutral absorption component to account for both Galactic and local extinction. 

High energy residual in a simple absorbed power-law clearly shows that there is a high energy roll over in the spectra which is similar to other ULXs. We henceforth, fitted the spectra with some phenomenological and physical models to explain the high energy turnover in the spectra.

The phenomenological model of a power-law with an exponential cutoff ({\texttt{cutoffpl}} in XSPEC) provides a significantly better fit compared to simple power-law fit for both epoch of observations ($\Delta \chi^2 = 158$ (Epoch 2) and $\Delta \chi^2 = 133$ (Epoch 3) for $1$ less degree of freedom). The folding energy is found to be $2.27^{+0.36}_{-0.28}$ keV and $3.86^{+0.68}_{-0.51}$ keV for Epoch 2 and Epoch 3  respectively. Further, a multicolour disk blackbody ({\texttt{diskbb}} in XSPEC) model is fitted to find the contribution of thermal disk component in the spectra. We also used the ``slim disk" (\texttt{diskpbb} in XSPEC) model for the ULX1 spectra in both epochs and found that  this model is statistically preferred over the hot Shakura \& Sunayev keplerian thin disk. The best fit values for $p$ in table ~\ref{tab:param1} clearly indicates that the disk emission is super-Eddington in nature (see section ~\ref{subsec:accretionstate} for details). The slim disk parameter, $p$, is $0.64^{+0.04}_{-0.03}$ and $0.61 \pm 0.02$ for Epoch 2 and Epoch 3, respectively. In both cases $p<0.75$ which proves the slim accretion disk geometry is preferred over thin accretion disk scenario.

\begin{figure}
	\includegraphics[width=\columnwidth]{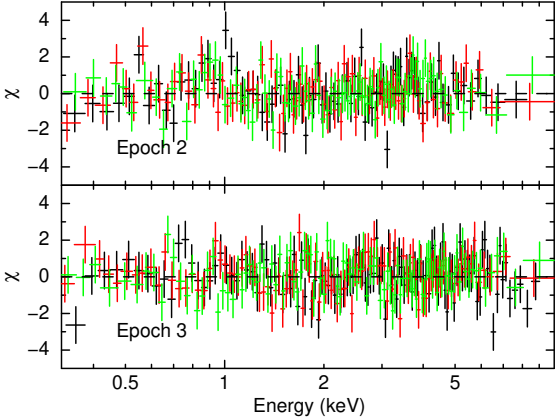}
    \caption{Residuals for disk blackbody with comptonization for two epochs of observation. Top panel is the residual for Epoch 2 analysis and bottom panel is for Epoch 3. Black, red and green represent PN, MOS1 and MOS2 spectra for both observations.}
    \label{fig:comptonized_corona_residuals}
\end{figure}

We also tried to fit the spectra with a thin multicolour disk (\texttt{diskbb}) and a Comptonization model (\texttt{comptt}) where seed photon temperature of the Comptonization is tied up with the inner disk temperature. Although we got statistically acceptable fit (see table ~\ref{tab:param2} for parameter values and figure ~\ref{fig:comptonized_corona_residuals} for residuals) for both observations, we found strong degeneracy between the Comptonization temperature and optical depth. The most likely reason for this is lack of hard X-ray coverage as \xmm\ does not go beyond about $10$ keV energy hence cannot constrain the Comptonization components. To determine the exact contribution of Comptonization process, we require high quality data above $10.0$ keV. We further tried to fit a slim accretion disk with comptonized corona but as before \xmm\ data was unable to properly constrain comptonization as well as disk parameters. Hence, we disregard this model completely for the time being until high energy data are available.  

Epoch 1 was maximally affected by particle flares and hence the corrected exposure for PN turns out to be only $103$ sec and about $4$ ksec for MOS1 and MOS2 each. Similar procedure as followed  in epochs 2 and 3 has been followed for this epoch 1 observation also. Model parameters are shown in the table ~\ref{tab:param1}. Here also, we found that the spectra have a cutoff (with folding energy $E_{fold} = 2.38^{+1.96}_{-0.62}$ keV)and give better fit than simple powerlaw ($\Delta \chi^2 = 14$ for $1$ less degree of freedom). The slim disk parameter in diskpbb model is $p = 0.65^{+0.14}_{-0.08}$. Within the low count statistics limit of this particular dataset, both the slim disk and thin disk model provide statistically similar fit (Table-\ref{tab:param1}). In fact for slim disk case, the $p$ value is not well constrained and the upper limit goes beyond $0.75$ which is the limit of  the ``slim disk" model. Due to lack of counts, we did not study comptonization corona for this observation.

\begin{table*}
	\centering
	\caption{Parameter table for different models in all three epochs of \xmm\ observation. Absorbed flux $F_x$ and luminosity $L_x$ is measured in $0.3-10.0$ keV energy range.}
	\label{tab:param1}
	\begin{tabular}{lcccr} 
		\hline
		Parameters & Unit & Epoch 1 & Epoch 2 & Epoch 3\\
		\hline
		 & & Model = TBabs*powerlaw & & \\
		\hline
		$N_H$ & $10^{22}$ cm $^{-2}$ & $0.22^{+0.06}_{-0.05}$& $0.28 \pm 0.02$ & $0.23 \pm 0.01$\\
		$\Gamma$ &  & $1.87 \pm 0.12$ & $1.98 \pm 0.04$ & $1.74 \pm 0.03$\\
		$N_{pl}$ & $10^{-4}$ & $5.55^{+0.76}_{-0.66}$ & $8.23^{+0.43}_{-0.41}$ & $11.45^{+0.42}_{-0.40}$\\
		$\chi^2$/dof & &$82/65$ & $410/260$ & $430/321$\\
		$F_x$ & $10^{-12}$ erg cm$^{-2}$ s$^{-1}$&$2.55^{+0.22}_{-0.20}$ & $3.24^{+0.11}_{-0.10}$ & $6.06 \pm 0.13$\\
		$L_x$ &$10^{+39}$ erg s$^{-1}$ &$3.42 \pm 0.28$ & $4.34^{+0.14}_{-0.12}$ & $8.10^{+0.18}_{-0.17}$\\
		\hline
		 & & Model = TBabs*cutoffpl& & \\
		\hline
		$N_H$ & $10^{22}$ cm $^{-2}$ & $ < 0.13$& $0.09 \pm 0.03$ & $0.11 \pm 0.02$\\
		$\Gamma$ &  & $0.61^{+0.56}_{-0.39}$ & $0.64 \pm 0.18$ & $0.93 \pm 0.12$\\
		$E_{fold}$ & keV &$2.38^{+1.96}_{-0.62}$ & $2.27^{+0.36}_{-0.28}$ & $3.86^{+0.68}_{-0.51}$ \\
		$N_{cpl}$ & $10^{-4}$ & $5.61^{+0.77}_{-0.67}$ & $8.32^{+0.43}_{-0.41}$ & $11.36^{+0.40}_{-0.39}$\\
		$\chi^2$/dof & & $68/64$ & $252/259$ & $297/320$\\
		$F_x$ & $10^{-12}$ erg cm$^{-2}$ s$^{-1}$ &$2.36^{+0.21}_{-0.20}$ & $3.05 \pm 0.10$ & $5.77^{+0.13}_{-0.14}$\\
		$L_x$ &$10^{+39}$ erg s$^{-1}$ &$3.16 \pm 0.27$ & $4.08 \pm 0.14$ & $7.71 \pm 0.18$\\
		\hline
		 & & Model = TBabs*diskbb & & \\
		\hline
		$N_H$ & $10^{22}$ cm $^{-2}$ & $ < 0.05$& $0.06 \pm 0.01$ & $0.05 \pm 0.01$\\
		$T_{in}$ & keV & $1.38^{+0.12}_{-0.11}$ & $1.31 \pm 0.04$ & $1.57 \pm 0.04$\\
		$N_{disk}$ & $10^{-2}$ & $3.13^{+1.14}_{-0.78}$ & $5.25^{+0.6}_{-0.5}$ & $4.67^{+0.44}_{-0.40}$\\
		$\chi^2$/dof & & $69/65$ & $265/260$ & $382/321$\\
		$F_x$ &$10^{-12}$ erg cm$^{-2}$ s$^{-1}$ &$2.32 \pm 0.19$ & $2.99 \pm 0.10$ & $5.55 \pm 0.13$\\
		$L_x$ &$10^{+39}$ erg s$^{-1}$ &$3.10^{+0.26}_{-0.25}$ & $4.00 \pm 0.13$ & $7.42 \pm 0.17$\\
		\hline
		& & Model = TBabs*diskpbb & & \\
		\hline
		$N_H$ & $10^{22}$ cm $^{-2}$ & $ < 0.16$ & $0.12 \pm 0.03$ & $0.14 \pm 0.02$\\
		$T_{in}$ & keV & $1.59^{+0.54}_{-0.30}$ & $1.52^{+0.12}_{-0.10}$ & $2.15^{+0.19}_{-0.15}$\\
		$p$ & & $0.65^{+0.14}_{-0.08}$ & $0.64^{+0.04}_{-0.03}$ & $0.61 \pm 0.02$ \\
		$N_{disk}$ & $10^{-2}$ & $ < 4.54 $  & $1.86^{+1.00}_{-0.67} $ & $0.75^{+0.34}_{-0.25} $\\
		$\chi^2$/dof & & $68/64$ & $248/259$ & $299/320$\\
		$F_x$ &$10^{-12}$ erg cm$^{-2}$ s$^{-1}$ &$2.33^{+0.21}_{-0.19}$ & $3.03 \pm 0.10$ & $5.73^{+0.13}_{-0.14}$\\
		$L_x$ &$10^{+39}$ erg s$^{-1}$ &$3.12^{+0.27}_{-0.26}$ & $4.05^{+0.14}_{-0.13}$ & $7.66 \pm 0.18$\\
		\hline
	\end{tabular}   
\end{table*}

\begin{table*}
	\centering
	\caption{MCD+comptonization model fitted for two epochs of \xmm\ observation. Epoch 1 was not used here due to low count statistics. Absorbed flux $F_x$ and luminosity $L_x$ is measured in $0.3-10.0$ keV energy range. $\dagger$ The parameter pegged at the low end and hence fixed to this value. }
	\label{tab:param2}
	\begin{tabular}{lccr} 
		\hline
		Parameters & Unit & Epoch 2 & Epoch 3\\
		\hline
		 & & Model = TBabs*(diskbb+comptt) & \\
		\hline
		$N_H$ & $10^{22}$ cm $^{-2}$ & $0.07 \pm 0.01$ & $0.08 \pm 0.01$\\
		$T_{in}$ & keV  & $0.97^{+0.26}_{-0.20}$ & $0.82^{+0.31}_{-0.26}$\\
		$N_{disk}$ &  & $0.13^{+0.12}_{-0.07}$ & $0.34^{+0.64}_{-0.21} $ \\
		$kT$ & keV & $2^\dagger$ & $<2.22$ \\
		$\tau $ & & $>5.65$ & $9.61^{+8.26}_{-1.46}$ \\
		$N_{comp} $ &$10^{-4}$ &$1.20^{+0.96}_{-1.00}$ & $3.67^{+2.25}_{-1.85} $ \\
		$\chi^2$/dof &  & $256/258$ & $300/318$\\
		$F_x$ & $10^{-12}$ erg cm$^{-2}$ s$^{-1}$ & $3.06^{+0.11}_{-0.10}$ & $5.78^{+0.12}_{-0.11}$\\
		$L_x$ &$10^{+39}$ erg s$^{-1}$  & $4.09 \pm 0.14$ & $7.72^{+0.17}_{-0.15}$\\
		\hline
	\end{tabular}
\end{table*}

\subsection{Variability}
\begin{figure}
	\includegraphics[width=\columnwidth]{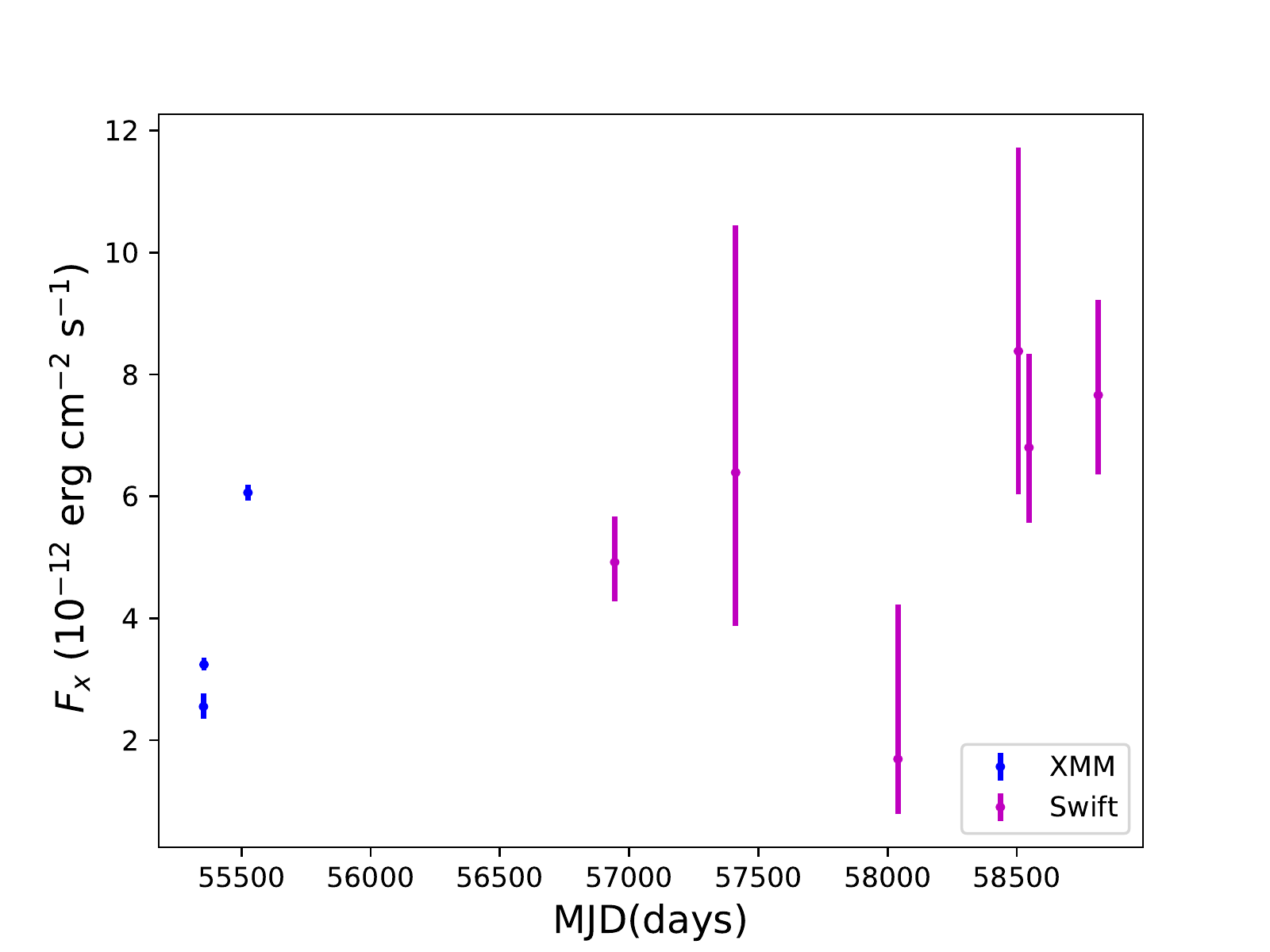}
    \caption{$0.3-10.0$ keV flux variation over time}
    \label{fig:flux_variation}
\end{figure}

\begin{figure}
	\includegraphics[width=\columnwidth]{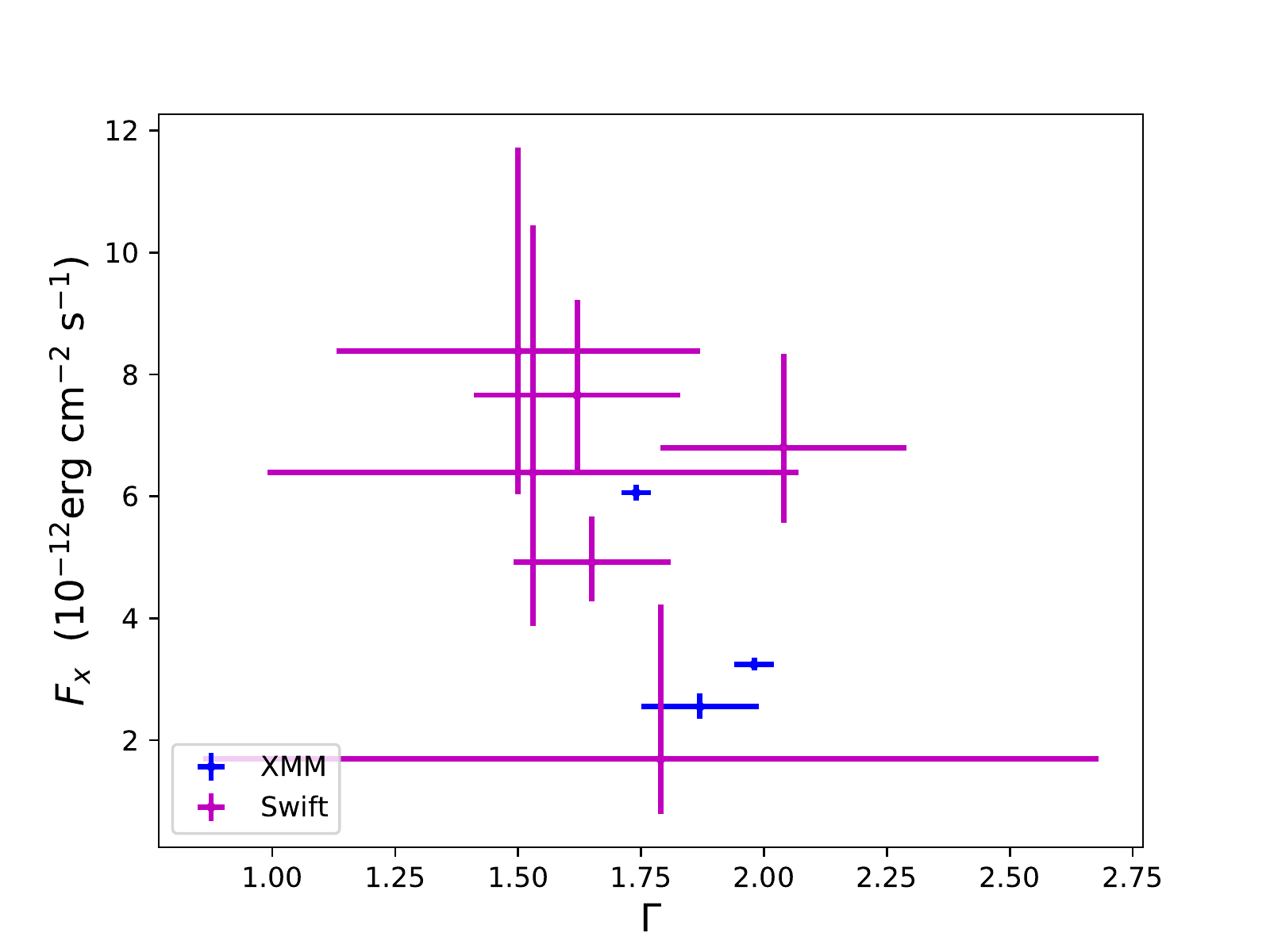}
    \caption{$0.3-10.0$ keV flux and spectral hardness relation. The source exhibits harder spectra with increasing flux.}
    \label{fig:F_G}
\end{figure}

NGC 4190 ULX1 is highly transient in nature. Although we do not find any short term variability, we have observed long term flux as well as spectral hardness variability. In order to study long term flux variability, we have included \swift\ data along with \xmm\ , even though they are only snap shot observations. Figure ~\ref{fig:flux_variation} shows the flux variation of the ULX over time as detected by \xmm\ and \swift\ . We report the absorbed flux and luminosity in $0.3-10.0$ keV energy range for both observatories throughout this paper unless mentioned otherwise. Since, the exposures for \swift\ -XRT are low, so is the signal to noise ratio, hence they provide relatively large errors in measurement of spectral parameters. Our analysis shows that with changing flux, there is a change in hardness of the spectra. In fact there seems to be a clear anti-correlation between flux and power-law photon index (see Figure ~\ref{fig:F_G}). To quantify this anti-correlation we used Pearson's ``r" correlation coefficient measurement technique and found the  correlation coefficient to be $-0.50$ with probability (``p" value) of $0.17$. That suggests the source is in spectrally harder state when brighter.  
 
\label{subsec:variable}

\section{Discussions}
\label{sec:discuss}

NGC 4190 ULX1 is a bright isolated source showing long term flux and spectral variability. Detail spectral analysis and variability study prove that this source is in an unusual accretion state compared to Galactic XRBs. 

Since we did not find any significant short term variability or pulsation in the source, we are unable to conclude whether this ULX hosts a neutron star (NS) or a black hole (BH). The high energy ($2-6$ keV) turnover in the spectra clearly indicates that a single power-law emission is not adequate to explain the X-ray emission process. This is a standard characteristic of ULXs, as most of the ULXs studied with broadband X-ray data show high energy turnover in the spectra \citep{Kaaret1, Bachetti1, Walton1, Rana1}.

\subsection{Accretion state of ULX1}
\label{subsec:accretionstate}
The \xmm\ spectra of NGC 4190 ULX1 are best represented with a modified (slim) disk model, and hence suggests that the accretion state of the source is best associated with the classification of ``broadened disk" with a ``curved" state \citep{Sutton1, Soria1}. In general the spectral curvature around $2-6$ keV in most of the ULXs is explained by various models which manifest ``ultraluminous state" of ULXs.  This is a consequence of the super-Eddington process which occurs when the accretion rate is near or few times above the standard accretion rate. The state is either very high state with a cool but optically thick comptonized corona or a modified inner disk dominated by radiation pressure, electron scattering, energy advection through radiation trapping and outflows (see \citet{Soria1} and references therein). Based on the observed curvature in the X-ray spectra of NGC~4190 ULX1, we can rule out the sub-Eddington {\it hard} canonical state of the source (\citet{Pintore1}). In order to see if the ULX is in sub-Eddington {\it soft} canonical state, a multicolour disk blackbody ({\texttt{diskbb}} in XSPEC) model is fitted to the spectra. We find that a slim disk geometry is preferred over a hot Shakura \& Sunayev keplerian thin disk in $0.3-10.0$ keV energy range. Therefore, we can conclude that the ULX is not in a sub-Eddington {\it soft} canonical state. Hence, the observed presence of curvature and slim disk geometry suggest that the source is not in canonical high/soft state. 

In the case when accretion rates of the disk are higher than Eddington limit, the state is known as super-Eddington state. In this case, outward radiation pressure increases the scale height of the innermost part of the disk and the advection becomes important. As a consequence, the radial temperature profile becomes; $T(r) \propto r^{-p}$, where $p$ is a free parameter which takes the value of $0.75$ in case of thin Keplerian disk. The preference of ``slim disk" model over the thin disk model in the spectra clearly shows that the disk emission is super-Eddington in nature. This suggests that the ULX contains a stellar mass compact object emitting X-rays with super-Eddington mechanism.

It is important to note that ``slim disk" model and a Keplerian disk with comptonized corona model give statistically acceptable fit in $0.3-10.0$ keV energy range. However, the comptonization parameters have physically unrealistic values (see table ~\ref{tab:param2}), as the comptonized up-scattered photons have a high energy excess at $\sim 20$ keV, and \xmm\ high energy cutoff is $\sim$10 keV. Therefore, we consider advection dominated disk as preferred model.

\subsection{Evolution of hardness-luminosity and temperature-luminosity relation in ULX1 }
NGC 4190 ULX1 is one of the very few ULX sources which has shown a clear anti-correlation between flux and power-law photon index. Similar characteristics have been observed in other ULXs like NGC 1313 ULX-2 (a PULX) and NGC 253 X-2, where the source becomes spectrally harder with increasing luminosity \citep{Kajava1}. 

\begin{figure}
	\includegraphics[width=\columnwidth]{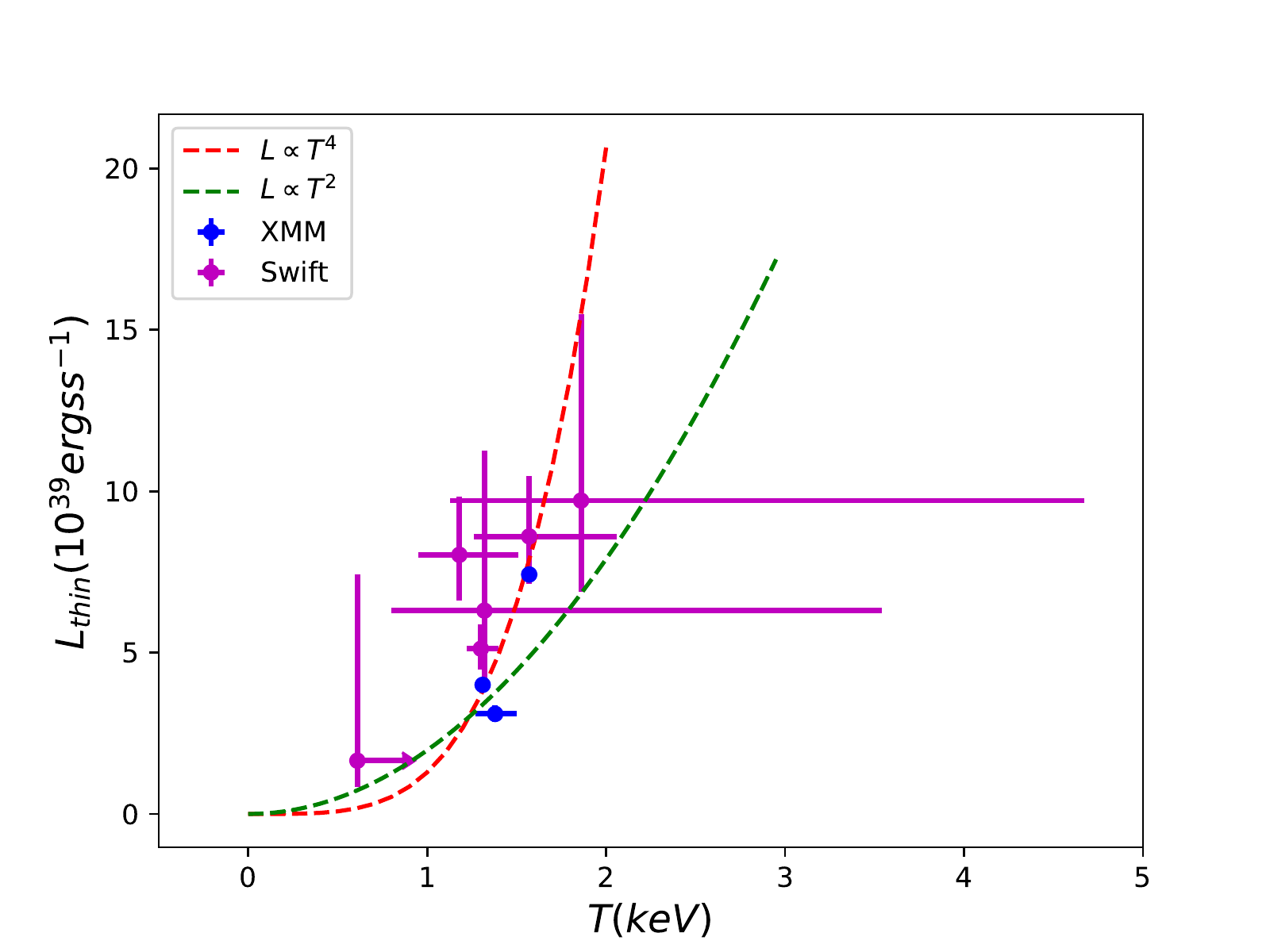}
    \caption{Luminosity-temperature relation for a thin accretion disk model. Red dashed line represents $L \propto T^4$ relation. Green dashed line represents $L \propto T^2$ relation. Thin disk model follows the $L \propto T^4$ relation and diverges away from $L \propto T^2$. }
    \label{fig:thinL_T}
\end{figure}

\begin{figure}
	\includegraphics[width=\columnwidth]{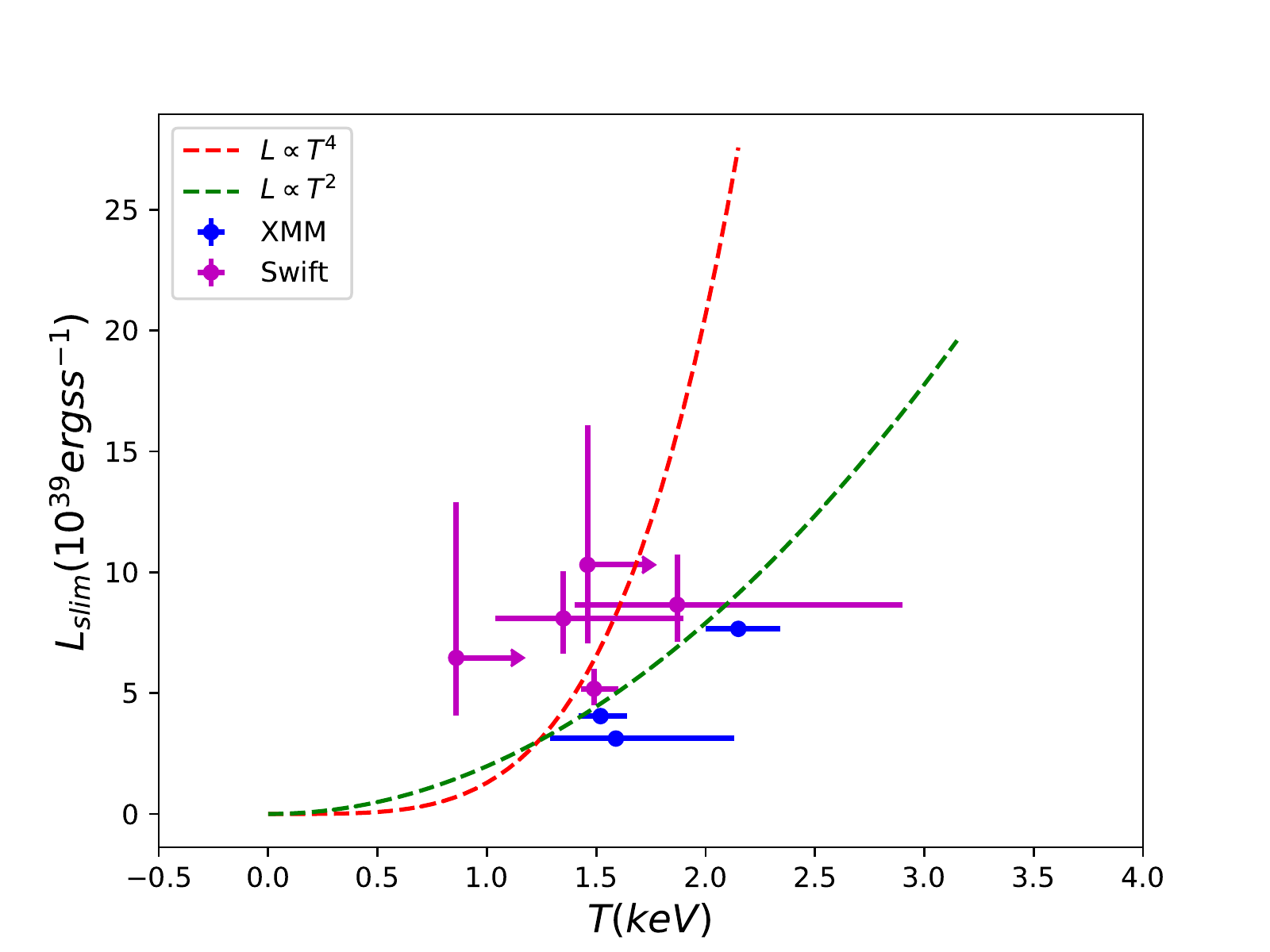}
    \caption{Luminosity-temperature relation for a slim accretion disk model. Colour scheme here is same as Figure \ref{fig:thinL_T}. Slim disk model apparently follows both $L \propto T^4$ and $L \propto T^2$ relations, however, if only good quality \xmm\ data is accepted, then it appears to be diverging from $L \propto T^4$ and seems to favour $L \propto T^2$ relation. One \swift\ observation could not constrain the fit and inner temperature pegged at high value $\sim 10$ keV, hence not shown here. }
    \label{fig:slimL_T}
\end{figure}

NGC 4190 ULX1 shows a positive Luminosity -temperature ($L-T$) relation in case of both thin disk and slim disk models. The $L-T$ plane of thin multi color disk model, follows $L \propto T^4$ relation (Fig. \ref{fig:thinL_T}) which is expected for a black body disk emission of a constant emitting area. The $L-T$ plane of slim disk model, whereas follows both $L \propto T^4$ and $L \propto T^2$ relation (Fig. \ref{fig:slimL_T}). However, $L \propto T^2$ relation is expected for an advection dominated disk \citep{Walton5} \citep{Kubota2} . It is important to note that the \swift\ data being unable to properly constrain the slim disk geometry because of its low count statistics, gives similar statistical confidence for both thin and slim disk model. Hence, if we disregard the \swift\ data in Fig. \ref{fig:slimL_T}, the good quality \xmm\ data shows a marginal preference towards the advection dominated accretion disk $L-T$ plane relation $L \propto T^2$ and can be seen diverging from $L \propto T^4$ relation.

\begin{figure*}
\centering

    \includegraphics[width=1.5\columnwidth]{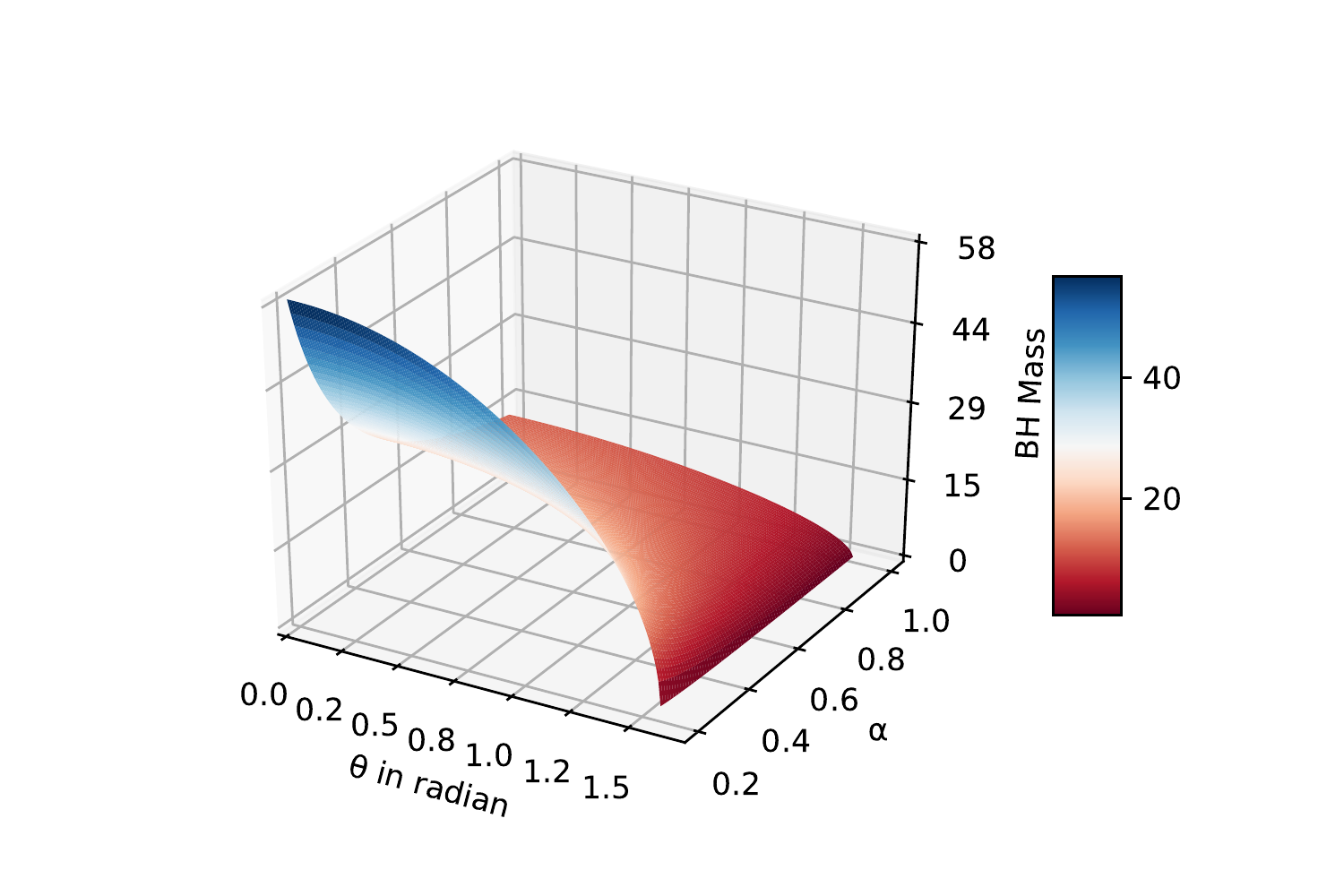}

    \caption{Black hole mass color map representation 3D diagram with varying inclination angle $\theta$ and $\alpha$ parameter.}
    \label{fig:BH_mass_representation}
\end{figure*}

\subsection{BH Mass Estimate}
\label{subsec:bhmass}

The spectral state of the source indicates a super-Eddington emission from a stellar mass compact object. In addition, we did not find any significant short term timing variability or pulsation in the \xmm\ data. To investigate from the spectral properties whether ULX1 hosts a NS or BH as the central compact object, we studied their spectral hardness and softness as prescribed by \citet{Pintore2017} for a best fit model of powerlaw with an exponential cutoff. We have calculated the hardness as the ratio of fluxes in $6.0-30.0$ keV and $4.0-6.0$ keV and softness as the ratio of fluxes in $2.0-4.0$ keV and $4.0-6.0$ keV. Fluxes beyond $10.0$ keV are calculated as an extrapolation of the $0.3-10.0$ keV best fit model. We found that its hardness and softness ratio falls in the range where most of the non-pulsating ULX systems reside \citep{Pintore2017}. Epoch 1 hardness and softness are $0.88 \pm 0.46$ and $1.95 \pm 0.35$ respectively. Epoch 2 hardness and softness are $0.79 \pm 0.18$ and $2.04 \pm 0.14$ respectively whereas epoch 3 hardness and softness are $1.63 \pm 0.17$ and $1.65 \pm 0.07$ respectively. In view of these calculations, we can expect that spectrally the ULX1 system manifests the nature of a black hole system. Based on this observed properties, we can safely assume that the central compact object is a black hole as typically considered for ULXs in general. Hence, we can estimate its mass given the source in all epochs have shown disk emission spectral characteristics. The physical inner radius $R_{in}$ can be determined from the disk normalization $N$ and hardening factor $\kappa$ which is the ratio of color temperature $T_{col}$ and effective temperature $T_{eff}$ and the geometric factor $\xi$ which appears due to the correction of apparent innermost radius $r_{in}$ from the physical innermost radius $R_{in}$, since the maximum disk temperature $T_{in}$ does not peak at $R_{in}$ \citep{Kubota1} \citep{Makishima1}. 

Since, 
\begin{equation}
       R_{in} = \xi \cdot \kappa^2 \cdot r_{in}
    \label{eq:two_radii}
\end{equation}

and the disk normalization relation is,

\begin{equation}
        N \approx (\frac{r_{in}}{D})^2 \cos\theta 
    \label{eq:norm}
\end{equation}

\begin{equation}
     R_{in} \approx \xi \kappa^2 N^{\frac{1}{2}} (\cos\theta)^{-\frac{1}{2}} D
    \label{eq:radius}
\end{equation}

where $R_{in}$ is in km and $D$ is in the units of $10$ kpc and $\theta$ is the inclination angle of disk.

Now, from the physical inner radius, we can estimate the mass for a black hole since this $R_{in}$ will be the Innermost stable circular orbit (ISCO) governed by the general relativistic gravitational potential.

\begin{equation}
     R_{in} = 3 \alpha R_s 
     \label{eq:mass}
\end{equation}

\begin{equation}
R_s = 2 \frac{G M}{c^2}
\end{equation}

where $M$ is the mass of the black hole, $c$ is the speed of light in vacuum, $R_s$ is defined as the Schwarzschild radius, $G$ is the gravitational constant and $\alpha$ is a function of spin parameter to take into account the most general spinning black hole scenarios.

Our analysis clearly shows that the NGC 4190 ULX1 is not in a canonical hard/soft state, hence we assumed the hardening factor and geometric correction factor as prescribed by \citet{Soria2, Watarai1, Vierdayanti1} taking $\kappa \approx 3$ since at higher accretion rates hardening factor increases and $\xi \approx 0.353$ which takes the transonic flow in the pseudo-Newtonian potential. 

Since, in epoch 1 observation, the data quality did not allow to constrain the value of normalization, we performed further calculations for epochs 2 and 3 observations.  In second epoch, the disk normalization is $N \approx 1.86^{+1.00}_{-0.67} \times 10^{-2} $ for diskpbb model and for the third epoch it is $N \approx 0.75^{+0.34}_{-0.25} \times 10^{-2} $. Using average normalization  value of $(0.87 \pm 0.28) \times 10^{-2}$ \citep{Barlow1} and equation ~\ref{eq:radius}, the physical inner radius turns out to be $R_{in} \approx 89^{+13}_{-16}$ km for a face on disk geometry. The face-on disk assumption provides the upper limit of the inner radius as well as the mass of the compact object.

Since for a ``slim" disk, the inner radius extends inside the ISCO, the true mass can be estimated from the ``apparent X-ray estimated"  mass as $M_{BH} \approx 1.2 M$ \citep{Vierdayanti1}. The spin parameter $\alpha$ takes different values for different physical scenarios. $\alpha = 1$ for non-rotating static and spherically symmetric Schwarzschild black hole, $\alpha = \frac{1}{6}$ for maximally rotating Kerr black hole and $\alpha = \frac{1.24}{6}$ for maximum possible spin achieved by an astrophysical black hole \citep{Thorne1}.

In the scenario of highest possible spin of a black hole, with the assumption of face on disk inclination, the average estimated mass of the black hole would be $M_{BH} \approx 58^{+9}_{-10} M_{\odot}$. Given that the system has a hot disk ($kT_{disk} > 0.5$ keV) with an average luminosity of $L_x \approx 5 \times 10^{39}$ \lumcgs, it is safe to consider that the black hole at the core of the ULX is a stellar mass black hole, which in a realistic case will consist of a mass $ 10 - 30 M_{\odot}$ \citep{Soria2, Vierdayanti1}.

The estimation of black hole mass on basis of the disk dominated spectral feature requires analysis of dependencies on rotation parameter $\alpha$ and the inclination angle of the disk with the line of sight $\theta$. Taking the average disk normalization, in figure \ref{fig:BH_mass_representation}, we show a 3D color map visual of how black hole mass ranges over different $\theta$ and $\alpha$ values, the only two free parameters in the mass estimation. It is clearly visible that even with highly rotating black hole and with small inclination angle, the mass will be $< 100 M_\odot$, which justifies our conclusion of NGC~4190 ULX1 to be a stellar mass black hole.

\section{Conclusions and Summary} 
Our detailed analysis of X-ray spectra from multiple \xmm\ observations suggest that ULX1 in NGC 4190 is not in standard canonical accretion state, normally observed in Galactic XRB sources. The 0.3-10.0 keV spectra shows a break at $E_{fold} \sim$ 2-4 keV which is a unique distinguishing feature of ULXs in ``ultraluminous state" when compared to Galactic XRBs.  The ``broadened disk state" with $p \sim 0.6$ clearly indicates the inner portion of the disk has a funnel like structure owing to the advection due to high outward radiation pressure. Long term variability study indicates that the source becomes spectrally harder with increasing X-ray flux. Unusual relation between X-ray flux and spectral slope (hardness), and the \xmm\ spectra being favoured by slim disk model proves that the source is in a super-Eddington state and hosts a stellar mass compact object. The slim disk luminosity-temperature ($L-T$) relation $L \propto T^2$, justifies the advective nature of the accretion flow in the inner part of the disk, since inner radius is inversely proportional to the inner temperature. From slim disk geometry as the best fit model within 0.3--10.0 keV energy range, we further estimated the mass to be maximum $\sim$ 10-30 $M_\odot$, thus a stellar mass compact object is the central power house in NGC~4190. Due to absence of any short term variability in time series, we are unable to conclude whether it is a stellar mass NS or BH. However, hardness and softness value of the spectra indicate that the host compact object is most likely a black hole. Therefore, in light of these data it is safe to state that, if this compact object is a black hole, it is a stellar mass black hole with mass of $ \sim 10-30$ \ms. Further investigation with broadband X-ray coverage and multi-wavelength study will be the key to obtain a clearer picture on the nature of the source and dominant physical mechanism at work in the source.

\label{sec:conclusion}

\section*{Acknowledgements}

We would like to thank referee for positive comments that helped in further improving the manuscript. This research has made use of archival data obtained with \textit{XMM-NEWTON}, an ESA science mission with instruments and contributions directly funded by ESA member states and NASA. This research has also made use of the archival data from \textit{Swift} observatory of NASA available at the High Energy Astrophysics Science Archive Research Center (HEASARC).

\section*{Data Availability}
The \textit{XMM-NEWTON} and \textit{Swift} data used for this work are all available for download from their respective public archives in High Energy Astrophysics Science Archive Research Center (HEASARC) .



\bibliographystyle{mnras}
\bibliography{example} 




\bsp	
\label{lastpage}
\end{document}